\documentclass[12pt,preprint]{aastex}

\usepackage{epsfig} 

\newcommand\etal{{et~al.}} 

\slugcomment{Accepted for publication in  ApJ}
\shorttitle{A Test on the Nonlinear Color-Metallicity Relation Scenario for Color Bimodality: M87 GCs in $u$-band}
\shortauthors{Yoon et al.}

\begin{document}

\title{Nonlinear Color-Metallicity Relations of Globular Clusters. II.\\
A Test on the Nonlinearity Scenario for Color Bimodality Using the $u$-band Colors: The Case of M87 (NGC 4486)}

\author{Suk-Jin Yoon\altaffilmark{1}, Sangmo T. Sohn\altaffilmark{2}, Sang-Yoon Lee\altaffilmark{1}, Hak-Sub Kim\altaffilmark{1}, Jaeil Cho\altaffilmark{1}, Chul Chung\altaffilmark{1}, and John P. Blakeslee\altaffilmark{3}}

\altaffiltext{1}{Department of Astronomy and Center for Galaxy Evolution Research, Yonsei University, Seoul 120-749, Korea}
\email{sjyoon@galaxy.yonsei.ac.kr}
\altaffiltext{2}{Space Telescope Science Institute (STScI), 3700 San Martin Drive, Baltimore, MD 21218}
\altaffiltext{3}{Herzberg Institute of Astrophysics, National Research Council of Canada, Victoria, BC V9E 2E7, Canada} 

\begin{abstract}
The optical color distributions of globular clusters (GCs) in most large elliptical galaxies are bimodal. Based on the assumed linear relationship between GC colors and their metallicities, the bimodality has been taken as evidence of two GC subsystems with different metallicities in each galaxy and led to a number of theories in the context of galaxy formation. More recent observations and modeling of GCs, however, suggest that the color-metallicity relations (CMRs) are inflected, and thus colors likely trace metallicities in a nonlinear manner. The nonlinearity could produce bimodal color distributions from a broad underlying metallicity spread, even if it is unimodal. Despite the far-reaching implications, whether CMRs are nonlinear and whether the nonlinearity indeed causes the color bimodality are still open questions. Given that the spectroscopic refinement of CMRs is still very challenging, we here propose a new photometric technique to probe the possible nonlinear nature of CMRs. 
In essence, a color distribution of GCs is a ``projected'' distribution of their metallicities.
Since the form of CMRs hinges on which color is used, 
the shape of color distributions varies depending significantly on the colors.
Among other optical colors, the $u$-band related colors (e.g., $u-g$ and $u-z$) are theoretically predicted to exhibit significantly less inflected CMRs than other preferred CMRs (e.g., for $g-z$). As a case study, we performed the {\it HST}/WFPC2 archival $u$-band photometry for the M87 (NGC 4486) GC system with confirmed color bimodality. We show that the $u$-band color distributions are significantly different from that of $g-z$, and consistent with our model predictions. With more $u$-band measurements, this method will support or rule out the nonlinear-CMR scenario for the origin of GC color bimodality with high confidence. The {\it HST}/WFC3 observations in F336W for nearby large elliptical galaxies are highly anticipated in this regard.
\end{abstract}
\keywords{galaxies: evolution --- galaxies: individual (M49, M60, M87) --- galaxies: star clusters --- globular clusters: general}
\section{INTRODUCTION}

Globular clusters (GCs) are among the oldest stellar systems in 
the observable universe and are always present in large galaxies.  
There is substantial evidence that GCs are the remnants of star 
formation events in galaxies, and are linked to the star formation, 
chemical enrichment, and assembly histories of their parent galaxies.  
Due to their bright and compact nature, GCs are detected 
out to great distances.  Moreover, each GC around a galaxy has a small 
internal dispersion in age and abundance\footnote{
We note that there is an increasing number of Galactic GCs with multiple stellar populations \cite[see, e.g.,][]{lee99,piotto02,norris04,bedin04,lee05b,yoon08,jwlee09,han09,gratton10}.
The GCs with internal age and/or abundance spreads tend to be at the bright end of the GC luminosity function (e.g., $
\omega$ Cen, M54, NGC 6388, and NGC 6441). More caution therefore should be taken regarding this issue when interpreting massive GCs in extragalaxies.}, 
and thus light from GCs is relatively easier to interpret than integrated light from complex stellar populations of galaxies.  
Such unique properties make 
them ideal objects for constraining the formation and evolution of 
their host galaxies (see West \etal\ 2004 and Brodie \& Strader 2006 for reviews).

One of the most significant discoveries in the field of extragalactic GCs 
over the past few decades is the bimodal optical color distributions 
(e.g., $C-T_1$, $V-I$, and $g-z$) of GC systems. The first recognition and 
statistical study of GC color bimodality was done by \citet{zepf93}, 
and the succeeding observations have found 
that this feature is ubiquitous in a majority of massive galaxies
\citep{ostrov93,whitmore95,mglee98,gebhardt99,harris01,kundu01,larsen01,peng04a,peng04b,peng06,harris06,jordan09,sinnott10,liu11}.

Because GCs in the Milky Way and other galaxies are usually older than 10 Gyr 
and age does not strongly affect broadband colors of GCs this old, 
their colors are primarily governed by the metallicity.
Empirical relations between the most often used broadband color, $V-I$, 
and [Fe/H] have been traditionally fit with a linear function. 
Using the more metallicity-sensitive colors $C-T_{1}$ or $C-R$, 
Harris \& Harris (2002) and Cohen, Blakeslee, \& C\^{o}t\'{e} 
(2003) found a mildly quadratic or broken linear relation 
between color and [Fe/H].
With the linear or mildly curved 
color-to-metallicity conversion, the bimodality observed in GC 
color distributions has been widely interpreted as bimodal 
metallicity distributions and hence taken as evidence of two distinct GC subsystems.
The physical origin of two GC subgroups within a single galaxy 
and its implications in the context of galaxy formation have been 
the topics of much interest \cite[e.g.,][]{ashman92,forbes97,cote98,mglee10a}.

The key assumption behind the interpretation that bimodal 
color corresponds to bimodal metallicity is the {\it linear} relationship 
between intrinsic metallicities and, their proxies, colors. 
Indeed, simple linear conversion of photometric colors to metallicities 
is a reasonable first-order assumption for obtaining the mean values of metallicities for GC systems.
But for investigating the detailed structure of GC metallicity distributions, including possible
two metallicity groups, the form of the color-metallicity relation (CMRs) must be known to higher
order.  In general, the slope of the dependence of a given photometric color on the
logarithmic metallicity [Fe/H] will change as a function of metallicity.
It has been known for decades that the color of the giant branch in Galactic GCs is a
nonlinear function of [Fe/H] \cite[e.g.,][]{michel84}.
Recent observations \citep{peng06,mglee08} 
and theories (Lee, Lee, \& Gibson 2002; Yoon, Yi, \& Lee 2006; Cantiello \& Blakeslee 2007) have found departures from linearity for integrated GC colors as a function of [Fe/H].
If the {\it bona-fide} shape of CMRs
is nonlinear, what has been thought to be metallicity distribution functions of GC systems 
may deviate significantly from the true distributions.
Knowing their precise CMR forms therefore will be crucial 
to our understanding of the formation of GC systems and their host galaxies.

In this paper, given that the direct spectroscopic refinement of CMRs is still very challenging, 
we propose a new photometric tool to probe the nonlinear nature of CMRs:
the photometric color distributions involving the $u$ bandpass (or ``u-band color distributions''). 
Section 2 recapitulates the nonlinear-CMR scenario for the origin of GC color bimodality as proposed in Yoon, Yi, \& Lee (2006, hereafter Paper I). 
Section 3 presents a case study of the M87 (NGC 4486) GC system with confirmed color bimodality,
and shows that the {\it HST}/WFPC2 F336W ($u$-band) colors  (i.e., $u-g$ and $u-z$) are remarkably consistent with the nonlinear-CMR hypothesis.
Thanks to the advent of the {\it HST}/WFC3, deep F336W photometry of nearby elliptical galaxies 
will be available in the near future. In this regard, section 4 discusses our theoretical predictions targeting at the {\it HST}/WFC3 F336W-band color distributions of two well-known Virgo giant elliptical galaxies (M49 and M60) with confirmed, strong color bimodality. 

\section{The Nonlinear Color-Metallicity Relation Scenario for Globular Cluster Color Bimodality}

The possibility of inflected, nonlinear CMRs has been extensively 
investigated on both observational and theoretical grounds. 
Peng \etal\ (2006) presented an empirical relation between 
color and metallicity using observed $g-z$ colors and 
spectroscopic measurements of [Fe/H] for GCs in the Milky 
Way, M49 and M87. 
The observed relation is tight enough to show a notable departure 
from linearity. 
The relation between [Fe/H] and 
$g-z$ is steep for [Fe/H] $<$ $-$0.8, shallow up to [Fe/H] 
$\simeq$ $-$0.5, and then steep again at higher metallicities.  
They proposed a piecewise linear relation broken at $g-z$ $\simeq$ 1.05 or [Fe/H] $\simeq$ $-0.8$. 
Independently, Paper I presented a theoretical metallicity-color relationship that has a significant inflection and reproduces well the observed $g-z$ CMR by Peng \etal\ (2006). The nonlinear nature of the relation between intrinsic metallicity and its proxy, colors, may hold the key to understanding the color bimodality phenomenon. 
Paper I showed that the wavy feature projects equidistant metallicity intervals near the quasi-inflection point 
(i.e., the most metallicity-sensitive point) onto larger color intervals, 
and thus can produce bimodal GC color distributions from a broad underlying [Fe/H] distribution, 
even if it is unimodal.
The nonlinear-CMR scenario gives a simple and cohesive explanation for the key observations, including ($a$) the overall shape of color histograms, ($b$) the number ratio of blue and red GCs as a function of host galaxy luminosity, and ($c$) the peak colors of {\it both} blue and red GCs as a function of host luminosity.

The observational and theoretical evidence for and against the nonlinearity of the CMRs, 
as well as the alternative hypothesis of bimodality in the intrinsic metallicity distributions, 
has been addressed in detail by Blakeslee et al. (2010) and Yoon et al. (2011, hereafter Paper III).
We refer the reader to those works for further discussion.

Figures 1 and 2 convey the essence of the Paper I explanation.
Figure 1 shows the synthetic color-magnitude diagrams for individual stars of the model GCs, 
and compares the resulting theoretical CMRs with observations. 
The stellar population simulations are based on the Yonsei Evolutionary Population Synthesis (YEPS) model\footnote{The models in this study are constructed using the Yonsei Evolutionary Population Synthesis (YEPS) code. The YEPS model generates ($a$) synthetic color-magnitude diagrams for individual stars \cite[see, e.g.,][]{lee94,lee99,lee05b,rey01,yoon02,yoon08,han09} and ($b$) synthetic integrated spectra for colors and absorption indices of simple and composite stellar populations \cite[see, e.g.,][]{lee05a,par97,lee00,rey05,rey07,rey09,kaviraj05,kaviraj07a,kaviraj07b,kaviraj07c,ree07,yoon06,yyl09,yoon09,spitler08,mieske08,choi09,cho11,yoon11}. One of the main assets of our model is the consideration of the systematic variation in the mean color of horizontal-branch stars as functions of metallicity, age, and abundance mixture of stellar populations. The standard YEPS model employs the Yonsei-Yale (Y$^2$) stellar evolution models (Y. Kim \etal\ 2002; Han \etal\ 2011, in prep.) and the BaSeL flux library (Westera \etal\ 2002). The spectro-photometric model data of the entire parameter space are available at http://web.yonsei.ac.kr/cosmic/data/YEPS.htm.} (Chung et al. 2011; Yoon et al. 2011, {\it in prep.}).
The upper left quadrant presents the synthetic Log {$T_{\it eff}$} vs. Log $L/L_{\sun}$ diagrams for 14 Gyr GCs with selected [Fe/H]'s, at which the departure of CMRs from linearity is most prominent. The selected [Fe/H]'s are marked by horizontal arrows in the CMRs in the other three quadrants. 
The modeled and observed CMRs for $g-z$ (upper right quadrant), $u-z$ (lower left), and $u-g$ (lower right) are displayed along with the color-magnitude diagrams generated from the synthetic Log {$T_{\it eff}$} vs. Log $L/L_{\sun}$ diagrams in the upper left quadrant. 
Table 1 summarizes the references to the observed data used in the CMRs.
Tables 2, 3, and 4 give the theoretical $g-z$, $u-z$, and $u-g$ CMRs, respectively, based on the YEPS model.

This section focuses on the $g-z$ CMR (upper right quadrant). An inspection of the observed $g-z$ CMR suggests that it follows an inverted S-shaped ``wavy'' curve with a quasi-inflection point at $g-z$ $\simeq$ $1.3$.
The YEPS model suggests that the observed wavy feature in the $g-z$ CMR is a consequence of two complementary effects: ($a$) The integrated color of the stars on the main-sequence and red-giant-branch stages (denoted by red isochrones in color-magnitude diagrams on the right) is a nonlinear function of metallicity at given ages. As a result, the CMR (dashed line) features a mild departure from linearity at lower metallicity. 
($b$) Standard stellar evolutionary theories predict nonlinear dependence of the mean color of horizontal-branch (HB) stars on metallicity \cite[e.g.,][]{ywlee90,yi97}. 
The accompanying color-magnitude diagrams on the right illustrate such effect of metallicity on the systematic HB color variation. The color of the HB changes at a brisk pace between [Fe/H] = $-0.5$ and $-0.9$ where the HB just departs from the red-clump position. The HB contribution further strengthens the departure from linearity for the $g-z$ CMR. 
As a combined effect of ($a$) the main-sequence and red-branch stars and ($b$) the HB stars, the $g-z$ color becomes several times more sensitive to metallicity between [Fe/H] = $-0.5$ and $-0.9$, resulting in a quasi-inflection point at [Fe/H] $\simeq$ $-0.7$. 
Note that, because the range of the rapid change in $g-z$ is as small as $\sim$ 0.5 in [Fe/H], the wavy feature would not be discernible in models with a [Fe/H] grid spacing larger than $\sim$ 0.3.


Deferring the examination on the $u-z$ and $u-g$ CMRs until the next section, 
we demonstrate in Figure 2 the effect of the nonlinear CMRs on observed color distributions of GCs. 
Our simulations target the M87 GC system, which exhibits a 
clear bimodality in the {\it HST} Advanced Camera for Survey (ACS) $g-z$ distribution (Peng \etal\ 2006)
and is one of few galaxies with deep {\it HST} $u$-band photometry (see \S3 below). 
The first column of Figure 2 shows how the inflected (or wavy) $g-z$ CMR causes bimodality in the $g-z$ color distribution.
The nonlinear feature in the CMR (top row) has the effect of projecting the equidistant [Fe/H] intervals onto broader $g-z$ intervals, causing scarcity at the quasi-inflection point on the CMR. As an aid to visualizing the simulated $g-z$ distribution, we plot the color vs. magnitude diagram in the second row. The divide between two vertical bands of GCs is immediately visible. The resulting $g-z$ histograms (third row) show clear bimodality: the scarcity near the quasi-inflection point is reflected as a clear dip.
The agreement between the simulated (third row) and 
observed (bottom row) $g-z$ distributions for the M87 GC system is remarkable,
without having to invoke a bimodal metallicity spread.

\section{A Critical Test for the Nonlinearity Scenario Using $u$-band Colors}

\subsection{Conversion from Metallicities to Colors}

Despite far-reaching implications of the nonlinear-CMR scenario for the origin of GC color bimodality (Paper I), 
whether CMRs are nonlinear and whether the nonlinearity indeed causes the color bimodality 
are still open questions.
Even the best samples currently available for the empirical color-metallicity calibrations 
are still relatively small and sparsely populated at the
high-metallicity end, and exhibit significant observational scatter
\cite[e.g.,][]{peng06,mglee08,beasley08,woodley11,alves11}.  
Larger samples of high-quality spectroscopic metallicities are needed to establish the
precise forms of the CMRs.

Given that the more stringent spectroscopic refinement of CMRs is still very challenging, 
we propose the $u$-band color distributions 
as a new tool to probe the nonlinear nature of CMRs. 
The nonlinearity issue drew our attention to the simple, basic fact that,
in essence, a color spread of GCs is a ``projected'' distribution 
of their metallicities. 
Since the form of CMRs hinges on which color is used, 
the shape of color distributions varies depending significantly on the colors. 
Hence a comparative analysis of the GC color distributions for different colors 
will offer a powerful tool to probe the presence of the nonlinear projection effect at work.
We have explored various combinations consisting of the 
commonly used broadband filters and found that among other optical colors, 
the $u$-band related colors (e.g., $u-g$ and $u-z$) are theoretically predicted 
to exhibit the most distinctive CMRs from other preferred CMRs (e.g., for $g-z$).

In order to examine the characteristic of the $u$-band colors, we turn back to Figure 1. 
The lower left and lower right quadrants of Figure 1 show 
the $u-z$ and $u-g$ CMRs, respectively.
For a given age, the CMRs for the $u$-band colors are substantially less 
inflected than the $g-z$ CMR. 
This is because the integrated $u$-band colors of main-sequence and red-giant-branch stars are smoother functions of metallicity compared to $g-z$.
In addition, the $u$-band colors are less sensitive 
to the temperature (i.e., Log {$T_{\it eff}$}) variation of HBs,
which is due to the fact that
the blueing effect of the optical spectra with increasing HB temperature 
is held back by the Balmer discontinuity where the $u$-band is located (Yi \etal\ 2004).
This is evidenced by individual HB stars in the color-magnitude diagrams (blue dots in the small panels on the right) of synthetic GCs with different metallicity.
Such properties of the $u$-band make the $u-z$ and $u-g$ colors good metallicity indicators for a wide range of age, 
and the $u$-band color distributions are expected to be
significantly different from distributions of other optical colors such as $g-z$, $V-I$, and $C-T_1$.
Comparison shows that $u-g$ is less inflected than $u-z$,
and, as a result, the degree of nonlinearity is in order of $g-z$, $u-z$, and $u-g$.

As a case study, we have selected M87 (NGC 4486), the central cD galaxy in the Virgo galaxy cluster. 
The M87 GC system is not only with confirmed color bimodality, but also
among few elliptical galaxies with deep $u$-band observations.  
We have downloaded WFPC2 F336W (hereafter, $u$) images
from the {\it HST} archive. 
The total exposure time is 28.8 ksec.
These data were used by 
Jord\'{a}n et al. (2002) for investigating the relative age difference 
between blue and red GCs in M87.
Our data reduction and photometry procedures are essentially
identical to those of Jord\'{a}n et al. (2002). In summary, we
processed the WFPC2 images with standard pipeline and
measured the brightness of each point source using the
DAOPHOT with aperture radii of 2 and 3 pixels for the
WF and PC chips, respectively. Aperture corrections were
derived and applied to a 0\farcs5 radius using bright GC
candidates, and charge transfer efficiency corrections were
applied following Whitmore, Heyer, \& Casertano (1999).
We then applied an additional correction of 0.1 mag to correct to infinite aperture
(Holtzman et al. 1995), and finally converted instrumental
mags to standard ABMAG system.
Our $u$-band catalog was matched with ACS/WFC $g$-
and $z$-band photometry of Jord\'{a}n \etal\ (2009) after
placing both the WPFC2 and the ACS catalogs to a
common coordinate system. A matching radius of 
1\arcsec\ was used. Sources were visually inspected to ensure that
the matching was done properly. 
Jord\'an et al (2009) selected {\it bona-fide} GCs with their mags, $g-z$ colors, and sizes. 
We further employed $u$-band color cuts ($u-g$ $<$ 0.8) to filter out contaminating sources, especially background
star-forming galaxies.
We compared our observed luminosity functions to those
presented in Jord\'{a}n et al. (2002) and did not find any
noticeable difference.

Figure 3 displays the observed GCs in M87 on the color-magnitude diagrams (top panels)
and on the color-color diagrams (bottom panels) along with their color distributions.
The ID, RA, DEC, $u$-, $g$-, and $z$-band mags, 
and their observational errors of $\sim$ 800 M87 GCs are given in Table 5.
In this study, we consider
$\sim$ 600 GCs (with $\sigma_u$ $<$ 0.2 mag) that have reliable $u$, $g$, and $z$ photometry in common,
and the sample is $u$-band limited.
In the color-magnitude diagrams (top panels), 
the divide between two vertical bands of GCs is immediately visible for $g-z$,
whereas the division appears less clear for $u-z$ and $u-g$.
In the color-color diagrams (bottom panels), 
as all the three colors get redder with increasing metallicity, 
they are directly proportional to one another.
The color-color relations are thus basically the metallicity sequences. 
The red loci represent our model predictions for coeval (13.9 Gyr) GCs 
from the metal-poorest ([Fe/H] = $-2.5$, top left point) to the metal-richest ([Fe/H] = 0.5, bottom right point).
The red crosses on each model line mark the uniform [Fe/H] intervals ($\Delta$[Fe/H] = 0.2 dex). 
The larger color intervals at the midpoint of each color-color relation
are consistent with the observed lower density of GCs at the intermediate metallicity. 
Despite the agreement between the observed and modeled color-color relations, 
there are slight offsets, which we attribute to the fact that current population simulations
are still incomplete in terms of the stellar evolutionary tracks and/or the atmospheric libraries.
The scatter around the color-color relations is mainly on account of the observational errors, 
although it may be partially due to possible spreads in the parameters of GCs,
such as age and [$\alpha$/Fe]. 
The grey histograms in bottom panels are the color distributions, 
which are used repeatedly in Figures 2, 4, and 5.

Back in Figure 2, the observed $u-z$ and $u-g$ color distributions of M87 GCs are shown 
in the bottom panels of the second and third columns, respectively. 
The observed $u$-band color distributions are systematically different from the $g-z$ distribution (first column), 
in that the prominence of bimodality found in $g-z$ is 
weakened in $u-z$ and diminished substantially in $u-g$. 
Although this is readily expected from the difference among the model CMRs in the degree of nonlinearity, 
one may wonder about the role of observational uncertainties in weakening bimodality. 
Table 6 shows that the typical photometric errors of both $u-z$ and $u-g$ are 
respectively 2.3 and 2.2 times larger than that of $g-z$,
but at the same time the ranges spanned by the colors 
are ($\Delta$($g-z$), $\Delta$($u-z$), $\Delta$($u-g$)) = (1.1 mag, 2.7 mag, 2.1 mag),
that is, the baselines of $u-z$ and $u-g$ are respectively 2.5 and 1.9 times longer than that of $g-z$.
As a result, the relative sizes of error bars are calculated to be ($g-z$ : $u-z$ : $u-g$) = (1.0 : 0.9 : 1.2). 
In a relative sense, the errors in the three colors are quite comparable to one another. 
Moreover, the $u-z$ color, which has {\it smallest} relative errors, still shows weakened bimodality in the color distribution.
It is, therefore, not likely that bimodality is simply blurred by larger observational errors in the $u$-band. 
In the third row, the observational uncertainties as a function of mag 
are fully taken into account in the simulated color distributions. 
The parameters that match up the morphologies of $g-z$, $u-z$, and $u-g$ color histograms {\it simultaneously} 
are 13.9 Gyr, $-0.5$ dex, and 0.6 dex for age, mean [Fe/H], and $\sigma$([Fe/H]), respectively.
Note that that the mean colors of modeled GCs are redder than the observation by 0.1 $\sim$ 0.2 mag.
Our stellar population models show that, for given input parameters, the {\it absolute} quantities of output are rather subject to the choice of stellar evolutionary tracts and model flux libraries, and the different choices can result in up to $\sim$ 0.2 mag variation in the $g-z$, $u-z$, and $u-g$ colors.
Hence, we put more weight on the {\it relative} color values of modeled GCs, i.e., the blue-to-red number ratios and the overall morphologies of the simulated color histograms.

We emphasize that, in the conventional view where 
GC colors linearly trace GC metallicities, there is no reason 
for the shape of color histograms to vary significantly with the colors in use
unless the observational uncertainties of different colors are incomparable. 
Two distinct GC subpopulations would manifest themselves even in different colors 
more or less in the same way. 
In contrast, in the context of the nonlinear 
CMRs, the variation in the histogram 
morphology for different colors is readily understood if the 
shape of the CMRs depends significantly 
on the choice of colors as evident from the top panels of Figure 2. 
Indeed, there is reasonable agreement between the theoretically predicted (third row) 
and observed (bottom) $u-z$ and $u-g$ color distributions for the case of M87 GCs.

\subsection{Inverse-conversion from Colors to Metallicities}

The above experiment suggests that a color distribution of a GC system does not directly 
expose its intrinsic metallicity distribution, but instead, for a given metallicity 
spread, the color distribution may be primarily determined by the 
exact form of the CMR. 
Motivated by the idea, we make a rather more ambitious attempt in Figure 4.
The metallicity-to-color conversion (\S\S\,3.1) should not be irreversible,
and we try to inverse-transform color distributions into metallicity distributions 
using the nonlinear CMRs. 
The inverse-conversion from colors to metallicities can be hampered by 
the different observational uncertainties depending on the colors of interest.
Moreover, the incompleteness of current models 
will be amplified by inverse-transformation, giving metallicity distributions in error.   
Nevertheless, a careful comparative analysis of the GC metallicity distributions 
that are independently obtained from different color histograms 
will shed some light on the color-metallicity nonlinearity hypothesis.

The inverse-transformations are applied again to the M87 GC system. 
The top panels of Figure 4 are identical to the bottom panels of Figure 2, 
and show the observed $g-z$, $u-z$, and $u-g$ distributions of the M87 GC system. 
The middle row shows that the $u$-band colors have significantly less inflected, ``smoother'' CMRs than $g-z$. 
Therefore, obtaining metallicity distributions from $u$-band color distributions via the $u$-band CMRs 
should be more straightforward than the case of $g-z$.   
The bottom row of Figure 4 presents the GC metallicity histograms derived from $g-z$, $u-z$, and $u-g$ colors 
that are based on the {\it inflected} color-to-metallicity conversions shown in the middle row. 
The best-fit parameters obtained in Figure 2 for the age and mean metallicity for M87 GCs are also used in Figure 4. 
The three observed color histograms with different morphologies (top row) are transformed 
into metallicity distributions (bottom row) that have a strong metal-rich peak with a metal-poor tail.
Although the broadness and the exact shape of wings of the inferred metallicity distributions 
are affected by the observational uncertainties in the colors of interest, 
the three metallicity histograms are remarkably consistent with one another 
in terms of their overall shape and their peak positions.
In contrast, the three distributions derived based on the conventional {\it linear} color-to-metallicity conversions 
(dotted lines in the middle row) are just replicas of their color histograms (dotted histograms in the bottom row), and thus do not agree with one another. 
We wish to emphasize that, under the nonlinear-CMR assumption, 
the typical shape of the metallicity distributions is obtained invariably from different colors, 
i.e., $g-z$, $u-z$, and $u-g$ for M87 GCs.
This occurrence indicates that the nonlinear CMR projection effect is at work for the colors examined in this study.

Figure 5 presents another experiment of the inverse-transformations. 
Following the conventional interpretations, 
the observed color distributions (top row) are divided into two distinct subgroups
and described as the sum (black lines) of two Gaussian distributions (blue and red lines). 
The bottom row gives the de-projected metallicity distributions of blue and red GCs 
that are based on the {\it inflected} color-to-metallicity conversions shown in the middle row. 
As one may expect from the result in Figure 4, 
the three different combinations of two Gaussian distributions (top row) 
are all transformed into metallicity distributions that are similar to one another (bottom row). 
Still, it may be possible that two subgroups do dwell even in a single unimodal metallicity distribution. 
However, it is interesting that the typical shape, characterized by a sharp peak with a metal-poor tail,
of the metallicity distributions seems to coincide with 
those produced by galaxy chemical evolution models assuming a virtually continuous chemical evolution through many successive rounds of star formation.
Perhaps more importantly, the typical shape is also similar to that of metallicity distributions for resolved field stars in nearby elliptical galaxies 
(e.g., Bird \etal\ 2010 for the M87 field-star metallicity distribution).
The implications of the typical shape of the inferred GC metallicity distributions
and its similarity to those from chemical evolution models and field-star observations 
are very important and sufficiently involved 
that we will discuss the issue in a separate paper (Paper III).

\section{Discussion and Conclusion}

In order to explore to what extent 
nonlinear metallicity-to-color transformations for GCs may be
responsible for turning a unimodal metallicity distribution into bimodal color distributions,
we have proposed a $u$-band color technique.
We showed that the addition of the $u$-band 
photometry to the existing $g$ and $z$ data has the potential of judging which 
CMRs are close to the true forms.
Using the {\it HST} photometry, we demonstrated that the $g-z$, $u-z$, $u-g$ color distributions 
for the rich GC system of M87 differ significantly, 
and all appear to be reasonably consistent with mapping 
from a single unimodal distribution of metallicity.
The results of our experiments on the M87 GCs in \S 2 and \S 3 strengthen 
the claim that the each CMR has a wavy form instead
of a linear one.  
Obviously, the next step is to carry out the test on the nonlinear-CMR scenario using more GC systems. 
In this section we discuss the theoretically predicted $u$-band color distributions of GCs in massive elliptical galaxies,  
anticipating that the $u$-band observations (e.g., using the {\it HST}/WFC3 F336W) 
of nearby massive elliptical galaxies will be available in the near future.

Our prediction of the $u$-band color distributions is made based on the existing observational data on the $g-z$ distributions using our theoretical CMRs.
Our simulations target 
two nearby giant elliptical galaxies, M49 and M60
in the Virgo galaxy cluster.
The galaxies meet the following criteria: 
($a$) the number of observed GCs is relatively large, 
($b$) deep $g$ and $z$ photometry exists in the {\it HST} archive, and 
($c$) their observed $g-z$ color distributions show clear bimodality. 
The two galaxies have ACS $g$ and $z$ photometry 
for more than 700 GCs \citep{jordan09}, and show strong $g-z$ color bimodality 
(Peng \etal\ 2006).
We note that the galaxies may have younger populations of GCs, 
but with the addition of the $u$-band measurements, identification of these 
younger GCs will be trivial (Hempel \& Kissler-Patig 2004; Yi \etal\ 2004; Kaviraj \etal\ 2005), 
thus allowing to use only the old GCs.

Figure 6 illustrates how the $u$-band 
observations can be used to improve our understanding 
of CMRs and metallicity distributions of GC systems.
We utilize existing $g$ and $z$ photometry for the galaxies (leftmost column). 
Note that, with the $u$-band limited samples, 
the red GCs in $g-z$ would be more subject to going under the detection limit, 
but the typical shape of the inferred metallicity distributions 
would persist as expected from the result of the M87 GC system. 
The right two columns show the predicted $u-g$ distributions under two different assumptions: 
the CASE 1 for the conventional linear CMRs (middle column); 
the CASE 2 for the nonlinear CMRs (rightmost column). 
The observed $g-z$ distribution of each GC system 
was first translated into the [Fe/H] domain (the vertical 
metallicity histograms along the y-axis) via the ($g-z$)-to-[Fe/H] 
(inverse) conversions. The hashed metallicity histograms (CASE 1) and the 
thick, blank metallicity histograms (CASE 2) are then converted to the $u-g$ 
distributions via the linear (CASE 1) and nonlinear (CASE 2) [Fe/H]-to-($u-g$) conversions. 
It is obvious that, bypassing obtaining the metallicity distributions, 
the direct conversions from $g-z$ to $u-g$ based on the color-to-color relations
would result in the identical $u-g$ distributions to those shown here. 
The resulting prediction of the $u-g$ distributions (filled 
histograms) shows that the two cases will be 
significantly different and easily distinguishable from each other.
The CASE 1 generates the $u-g$ distributions for M49 and M60 
that are almost identical to their $g-z$ distributions,
whereas the CASE 2 produces the $u-g$ distributions 
that are by far closer to broad, unimodal distributions than their $g-z$ distributions.

A given intrinsic metallicity distribution of GCs should be manifested by different color distributions 
depending on the passbands in use. 
We have shown the power of $u$-band in discriminating between the two 
competing scenarios for the form of CMRs. 
By the projection effect, any feature on CMRs is manifested on the color domain. 
Hence, under the assumption of the nonlinear CMRs, the $u$-band color distributions are significantly different and readily distinguishable from those under the assumption of the conventional linear CMRs. With more data, this method will support or rule out the nonlinear-CMR scenario for GC color bimodality with high confidence. 
Further $u$-band measurements for GC systems with color bimodality are clearly needed, and the {\it HST}/WFC3 observations in F336W for nearby large elliptical galaxies are highly anticipated in this regard.
It is also noteworthy that another path we can take is to extend the photometry to IR passbands
\cite[e.g.,][]{hempel07,kundu07,spitler08,chies10,chies11a,chies11b},
where the contribution from the HB is just as {\it systematic} as in the $u$ passband. 
Interestingly, the most recent NIR/optical photometric study 
\citep{chies10} finds that the GC bimodality of 14 early-type galaxies
behaves systematically in that it becomes less evident in $g-K_s$ and even weaker in $z-K_s$ 
when compared to $g-z$. 
If the nonlinearity of CMRs is found to be favored by future multiband observations involving the $u$- and IR-bands, 
it will change much of the current thought on the GC color bimodality 
as well as the formation of GCs and their host galaxies.

\clearpage
\acknowledgments 
We would like to thank the anonymous referee for useful comments and suggestions.
SJY acknowledges support from Mid-career Researcher Program (No. 2009-0080851) and Basic Science Research Program (No. 2009-0086824) through the National Research Foundation (NRF) of Korea grant funded by the Ministry of Education, Science and Technology (MEST), and support by the NRF of Korea to the Center for Galaxy Evolution Research and the Korea Astronomy and Space Science Institute Research Fund 2011.
SJY would like to thank Daniel Fabricant, Charles Alcock, Jay Strader, Dongsoo Kim, Jaesub Hong for their 
hospitality during his stay at Harvard-Smithsonian Center for Astrophysics as a Visiting Professor in 2011--2012.


\clearpage
\begin{figure*}
\begin{center}
\includegraphics[width=21cm, angle=90]{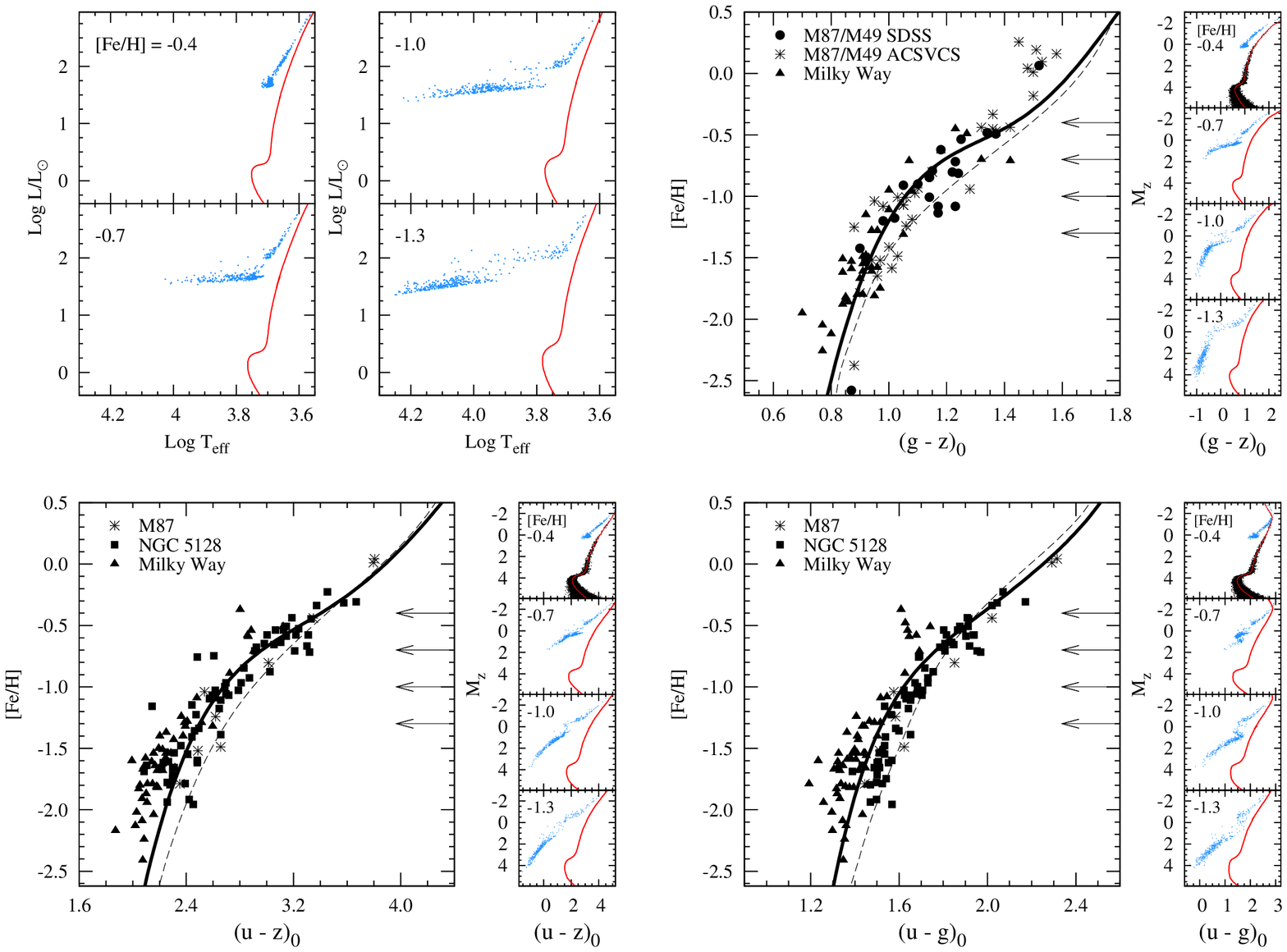}
\end{center}
\end{figure*}

\clearpage
\begin{figure*}
\begin{center}
\caption{\small
Empirical and theoretical CMRs
along with synthetic Log {$T_{\it eff}$}-Log $L/L_{\sun}$ and color-magnitude diagrams.
({\it Upper left quadrant}) 
Synthetic Log {$T_{\it eff}$} vs. Log $L/L_{\sun}$ diagrams of individual stars for 14 Gyr model GCs with various [Fe/H]'s.
Red loci are the Yonsei-Yale (Y$^2$) isochrones (Kim \etal\ 2002) depicting main sequence and red giant branch, 
whereas blue dots represent HB stars based on the Y$^2$ HB tracts (Han \etal\ 2011, in prep.).
({\it Upper right quadrant}) 
The observed relationship between $g-z$ and [Fe/H] 
for 22 M49 and M87 GCs with SDSS photometry (circles)
and 33 M49 and M87 GCs with ACS Virgo Cluster Survey photometry (asterisks), 
and 40 Galactic GCs with E($B-V$) $< 0.3$ (triangles). 
The references to the observed data and their selection criteria are summarized in Table 1. 
The thick solid line is for the 5th-order polynomial fit to our model prediction for 14-Gyr GCs,
and the dashed line is for the model without inclusion of the HB prescription. 
The model data for 10 $\sim$ 14 Gyr with fine grid spacing ($\Delta$[Fe/H] = 0.1) are given in Table 2.
The $\alpha$-element enhancement parameter, [$\alpha$/Fe], is assumed to be 0.3. 
Arrows denote the four values of [Fe/H], for which the synthetic color-magnitude 
diagrams are given in the right small panels. 
Synthetic color-magnitude diagrams on the right are generated from the synthetic Log {$T_{\it eff}$}-Log $L/L_{\sun}$ in the upper left quadrant
using the BaSeL flux library (Westera \etal\ 2002).
The top panel shows individual stars (black and blue dots) with an error simulation, 
whereas the rest panels show only the corresponding isochrones (red loci) and HB stars (blue dots).
({\it Lower left quadrant}) 
Same as the upper right quadrant, but for the $u-z$ color.
Asterisks, squares, and triangles represent GCs in M87, NGC 5128, and the Milky Way, respectively. 
The $u-z$ colors of the GCs in the Milky Way and NGC 5128 were converted from their $U-I$ colors via the equation, 
($u-z$) = 1.018 ($U-I$) + 0.607 (see Table 1), that were obtained from synthetic spectra of old (10 $\sim$ 15 Gyr) model GCs. 
({\it Lower right quadrant}) 
Same as the lower left quadrant, but for the $u-g$ color.
The $u-g$ colors of the GCs in the Milky Way and NGC 5128 were converted from their $U-B$ colors via the equation, 
($u-g$) = 1.014 ($U-B$) + 1.372 (see Table 1), that were obtained from synthetic spectra of old (10 $\sim$ 15 Gyr) model GCs. 
\label{fig1}}
\end{center}
\end{figure*}

\clearpage
\begin{figure*}
\begin{center}
\includegraphics[width=17cm]{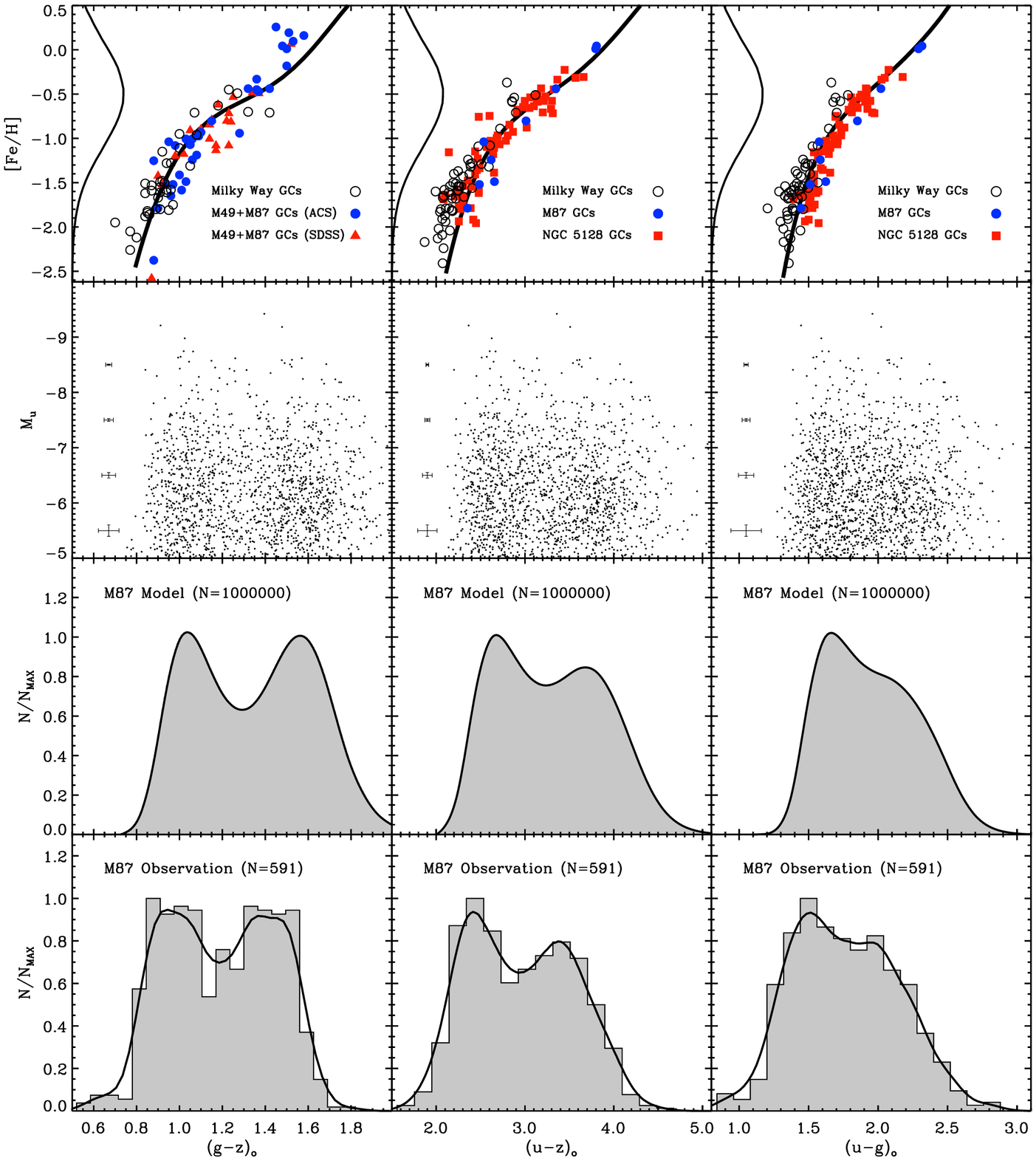}
\end{center}
\end{figure*}

\clearpage
\begin{figure*}
\begin{center}
\caption{
Mutiband ($u$, $g$, and $z$) observation of GCs in M87 
and Monte Carlo simulations for their color distributions. 
({\it Top row}) Same as the CMRs in Figure 1. 
The metallicity spread of 10$^6$ model GCs is shown along the y-axis, 
for which a simple Gaussian distribution is assumed 
($<$[Fe/H]$>$ = $-0.5$ dex and $\sigma$([Fe/H]) = 0.6).
The best-fit age to reproduce the morphologies of $g-z$, $u-z$, and $u-g$ color histograms {\it simultaneously} 
is 13.9 Gyr.
({\it Second row}) 
The left, middle, and right columns 
represent the color-magnitude diagrams of 2000 randomly selected model GCs
for the $g-z$, $u-z$, and $u-g$ colors, respectively.
The colors are transformed from [Fe/H]'s by using the theoretical relation shown in the top row.
For the integrated $u$-band absolute mag, $M_u$, 
a Gaussian luminosity distribution ($<$$M_u$$>$ = 25.2, distance modulus = 31.02, and $\sigma$($M_u$) = 1.15) 
is assumed according to the observation. 
Observational uncertainties as a function of $M_u$ shown by error bars are taken into account in the simulations.
({\it Third row}) 
The left, middle, and right columns 
represent the color distributions of 10$^6$ modeled GCs for the $g-z$, $u-z$, and $u-g$ 
colors, respectively. 
({\it Bottom row}) 
Same as the third row, but the observed color histograms for the M87 GC system (see Figure 3).
The 591 GCs were used that have $u$, $g$, and $z$ measurements in common, 
and the sample is $u$-band limited. 
Solid lines are smoothed histograms with Gaussian kernels of 
$\sigma$($g-z$) = 0.05, $\sigma$($u-z$) = 0.10, and  $\sigma$($u-g$) = 0.15, respectively.
\label{fig2}}
\end{center}
\end{figure*}

\clearpage
\begin{figure*}
\begin{center}
\includegraphics[width=22cm,angle=90]{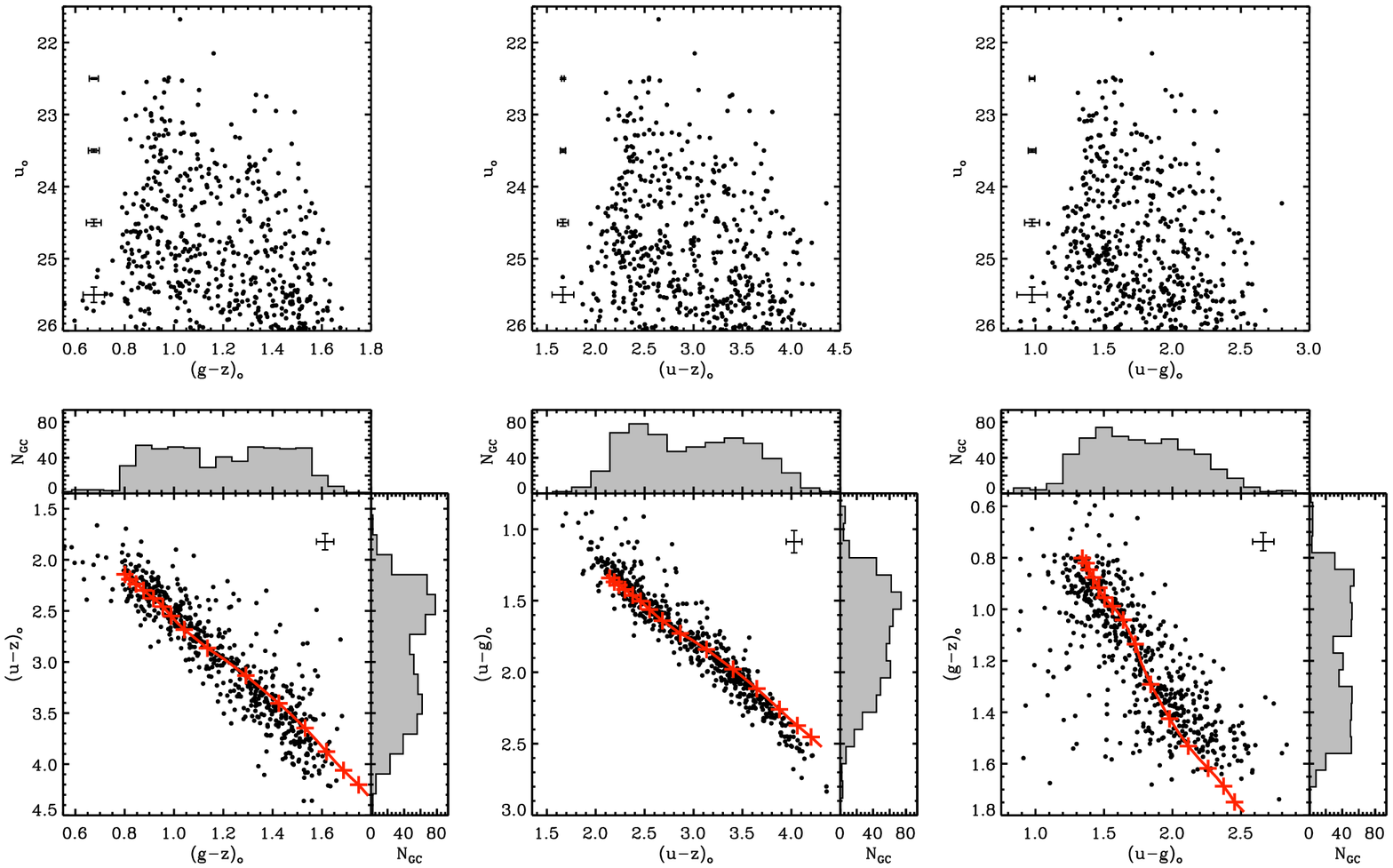}
\end{center}
\end{figure*}

\clearpage
\begin{figure*}
\begin{center}
\caption{
({\it Upper}) The color-magnitude diagrams for the GC system of M87. 
The left, middle, and right panels show the $g-z$, $u-z$, and $u-g$ distributions, respectively.
The $g$ (ABMAG) and $z$ (ABMAG) mags for GC candidates
are from the ACS GC catalog (Jord\'{a}n \etal\ 2009).
Using the {\it HST}/WFPC2 archival images, we measured 
$u$-band mags (F336W, ABMAG) for the M87 
GC candidates listed in the ACS GC catalog. 
A color cut ($u-g$ $<$ 0.8) were employed to filter out contaminating sources, presumably star-forming background galaxies.
({\it Lower}) The color-color diagrams and the projected color histograms for the GC system of M87. 
In this study, we used GCs that have $u$, $g$, and $z$ measurements in common, 
and the sample is $u$-band limited.
The $\sim$ 800 GC candidates are listed in Table 5.
Black dots in each panel are the selected $\sim$ 600 GCs ($\sigma_u$ $<$ 0.2 mag).
Red solid lines represent our model prediction for 13.9 Gyr GCs (Tables 2, 3, and 4)
from the metal-poorest ([Fe/H] = $-2.5$, top left point) to the metal-richest ([Fe/H] = 0.5, bottom right point).
The red crosses on each model line mark the uniform [Fe/H] intervals ($\Delta$[Fe/H] = 0.2 dex). 
\label{fig3}}
\end{center}
\end{figure*}

\clearpage
\begin{figure*}
\begin{center}
\includegraphics[width=17cm]{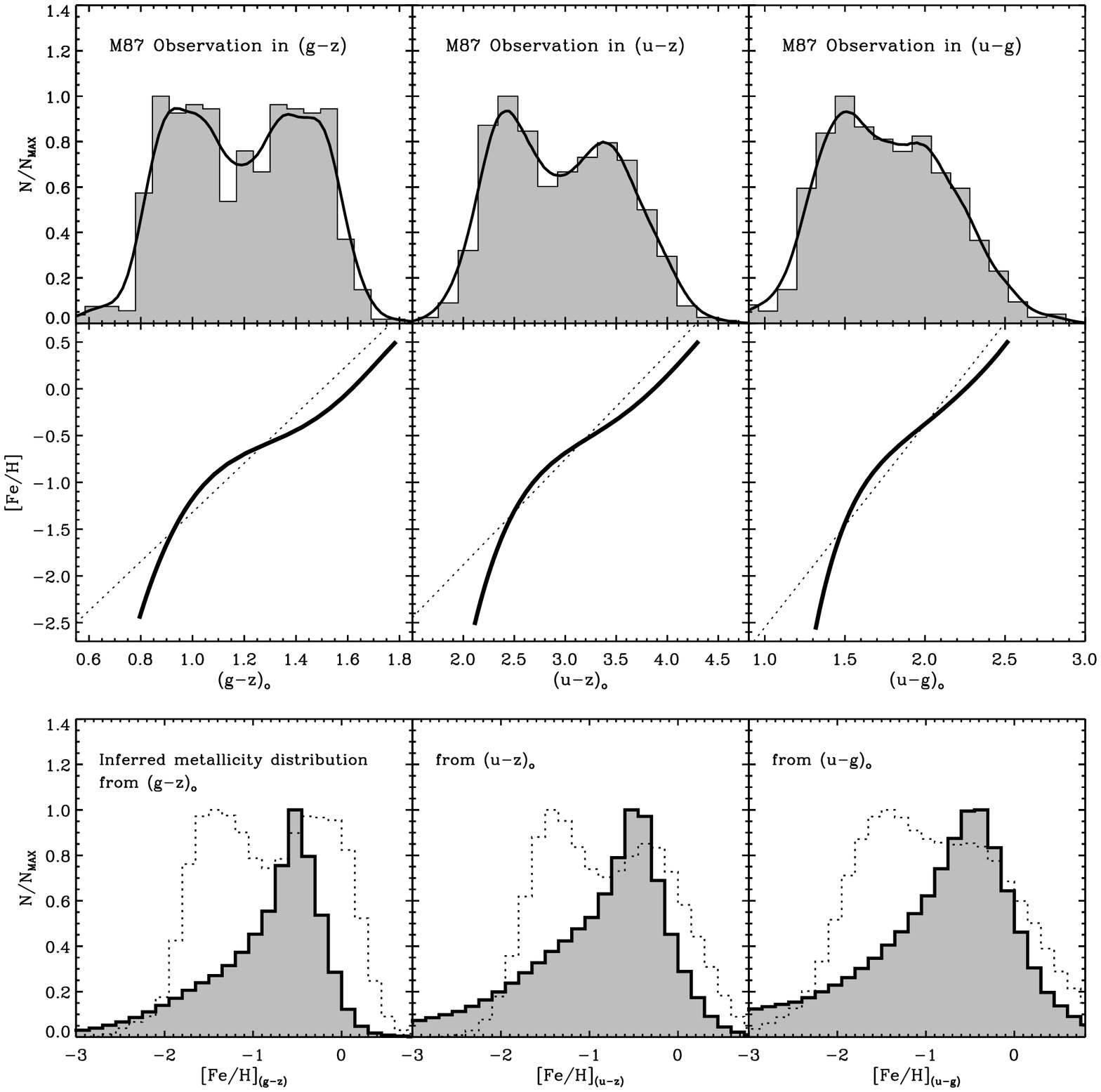}
\end{center}
\end{figure*}

\clearpage
\begin{figure*}
\begin{center}
\caption{
The $g-z$, $u-z$, and $u-g$ color distributions and their inferred metallicity distributions for the GC system of M87.
({\it Top row}) Same as the bottom row of Figure 2.
({\it Middle row}) Same as the top row of Figure 2, 
but the dotted lines represent the least-squares fits to the observational data points 
which are not shown here for clarity. 
({\it Bottom row}) 
The left, middle, and right panels show GC metallicity distributions 
obtained from the $g-z$, $u-z$, and $u-g$ color distributions, respectively.
The metallicity distributions are derived from the smoothed histograms in the top row 
via the nonlinear transformations (grey histograms with solid lines)
and via the linear transformation (open histograms with dotted lines). 
\label{fig4}}
\end{center}
\end{figure*}

\clearpage
\begin{figure*}
\begin{center}
\includegraphics[width=17cm]{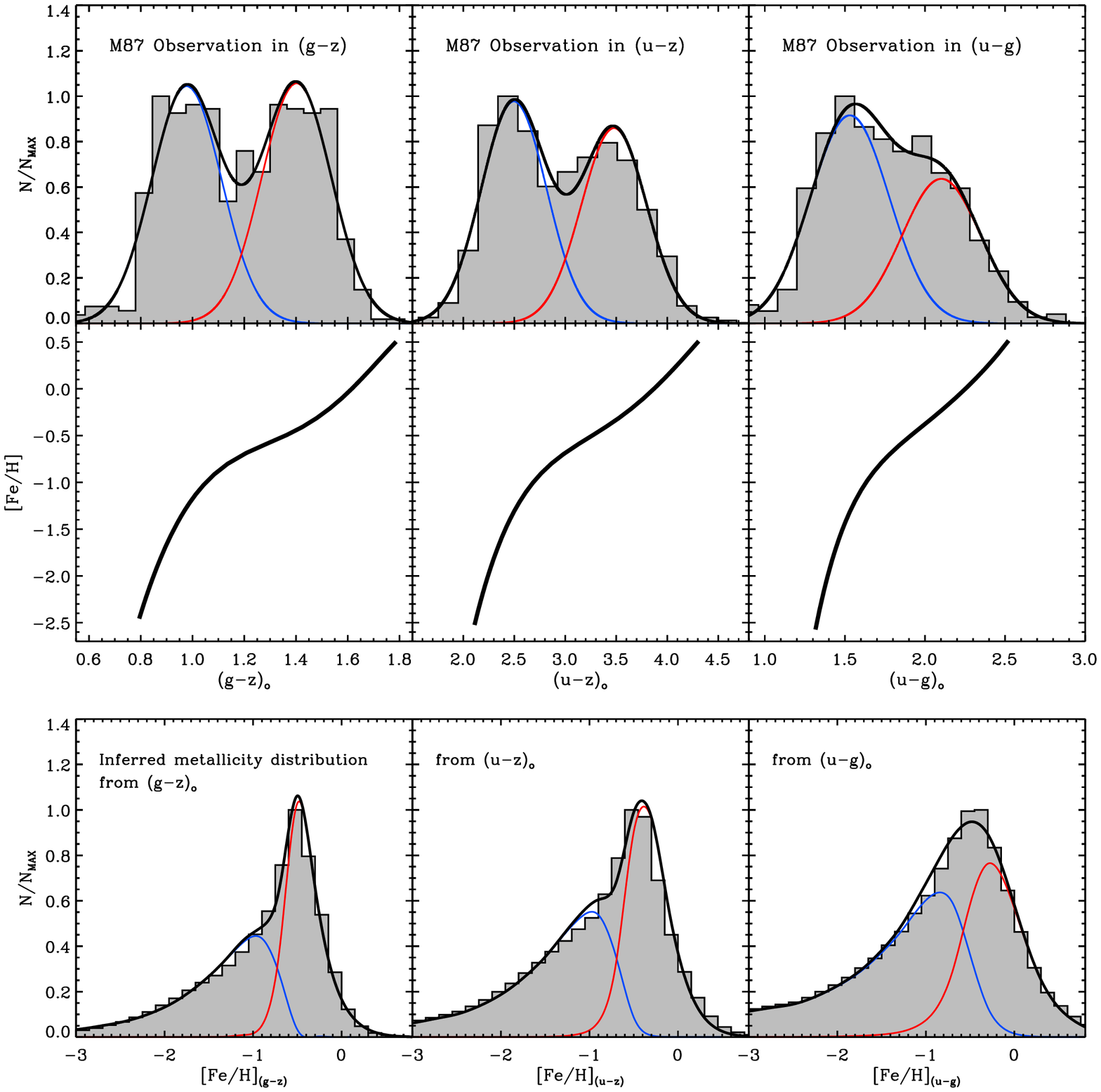}
\end{center}
\end{figure*}

\clearpage
\begin{figure*}
\begin{center}
\caption{
Same as Figure 4, but the observed histograms are expressed by a sum of two unimodal distributions.
({\it Top row}) Same as the top row of Figure 4, 
but the histograms are expressed by two (i.e., blue and red) Gaussian distributions based on the KMM analysis.
The blue, red, and black lines represent the blue, red, and total GCs,
with the peak colors and number fractions of blue and red GCs being
[(0.98, 1.40), (50 \%, 50 \%)] for $g-z$, 
[(2.50, 3.48), (53 \%, 47 \%)] for $u-z$, 
and [(1.53, 2.10), (59 \%, 41 \%)] for $u-g$, respectively. 
({\it Middle row}) Same as the middle row Figure 4, 
but without the observational data or the linear least-squares fits. 
({\it Bottom row}) Same as Figure 4, 
but the GC metallicity distributions are derived from the blue and red Gaussian distributions in the top row
through the color-to-metallicity transformations in the middle row. 
The blue, red, and black lines represent the blue, red, and total GCs.  
\label{fig5}}
\end{center}
\end{figure*}

\clearpage
\begin{figure*}
\begin{center}
\includegraphics[width=17cm]{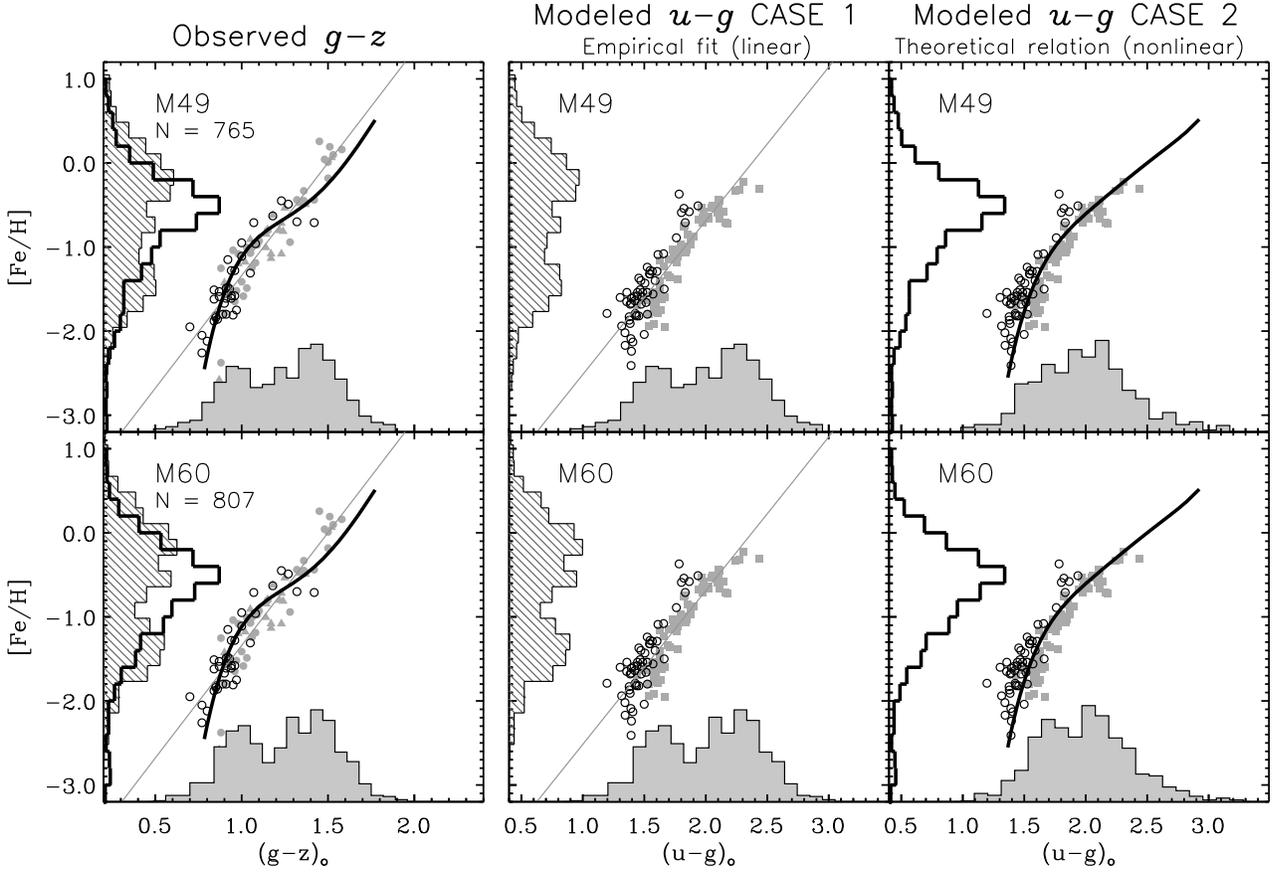}
\caption{
Observed $g-z$ distributions (leftmost column, Jord\'{a}n \etal\ 2009) and predicted $u-g$ distributions (middle and right columns) 
for the GC systems in our target galaxies, M49 (upper row) and M60 (lower row).   
The predicted $u-g$ distributions of each GC system are obtained under the two differing assumptions: 
Case 1 for the empirical linear CMRs (from least-squares fit to the observed data) 
and Case 2 for the theoretical nonlinear CMRs (from the YEPS model). 
The 13.5 Gyr is assumed for the theoretical CMRs.
The observed $g-z$ distribution (leftmost column) of each GC system was first translated into the [Fe/H] domain 
(the vertical metallicity histograms along the y-axis) via the linear (Case 1) and nonlinear (Case 2) $g-z$ vs. [Fe/H] relations. 
The hashed metallicity histograms (Case 1) and the empty metallicity histograms with thick solid lines (Case 2) 
are then converted to the $u-g$ distributions (the horizontal metallicity histograms along the y-axis) 
via the linear (Case 1) and nonlinear (Case 2) [Fe/H]-to-($u-g$) conversions. 
\label{fig6}}
\end{center}
\end{figure*}


\clearpage
\begin{table}
\footnotesize
\begin{center}
\caption{References to the observational data for the [Fe/H] vs. color relations used in this study.\label{tbl-1}}
\vspace{0.5cm}
\begin{tabular}{llll}
\tableline
\tableline
{Relations}  &  Galaxy Name & \multicolumn{2}{c}{References and Selection Criteria} \\
\tableline
			      		 		&  				& Spectroscopic [Fe/H] 				& Broadband Color 							\\
\tableline
The [Fe/H] vs. $g-z$ relation		& Milky Way 		& 1, 2 							& 1, 2 									\\
							& M49	 		& 1, 2 							& 1, 2 									\\
							& M87 			& 1, 2 							& 1, 2 									\\
\tableline
The [Fe/H] vs. $u-z$ relation  	& Milky Way 		& 3 								& 3 [($U-I$)\tablenotemark{a}, E($B-V$) $<$ 0.3]	\\
($u$ = {\it HST}/WFPC2 $F336W$)	& NGC 5128		& 4, 5 [$t$ $>$ 8 Gyr, S/N $>$ 10]	& 6 [($U-I$)\tablenotemark{a}]					\\
							& M87  			& 1, 2 							& This study [($u-z$)] 						\\
\tableline
The [Fe/H] vs. $u-g$ relation		& Milky Way 		& 3 								& 3 [($U-B$)\tablenotemark{b}, E($B-V$) $<$ 0.3]	\\
($u$ = {\it HST}/WFPC2 $F336W$)	& NGC 5128		& 4, 5 [$t$ $>$ 8 Gyr, S/N $>$ 10]	& 6 [($U-B$)\tablenotemark{b}]					\\
 							& M87  			& 1, 2 							& This study [($u-g$)] 						\\
\tableline
\tablenotetext{a}{The equation, ($u-z$) = 1.018 ($U-I$) + 0.607, is used to obtain $u-z$ from $U-I$.}
\tablenotetext{b}{The equation, ($u-g$) = 1.014 ($U-B$) + 1.372, is used to obtain $u-g$ from $U-B$.}
\end{tabular}
\tablerefs{(1) \citet{peng06}; (2) Paper I; (3) Harris \etal\ (1996, the 2010 edition); (4) \citet{beasley08}; (5) \citet{chung11}; (6) \citet{peng04a,peng04b}.}
\end{center}
\end{table}

\clearpage
\begin{rotate}
\begin{table}
\tiny
\caption{\footnotesize Theoretical $g-z$ colors for synthetic GCs with various ages ($t$) based on the YEPS models without ($w/o$) and with ($w$) the inclusion of the HB prescriptions (the 5th-order polynomial fits to the model data).\label{tbl-2}}
\begin{tabular}{ccccccccccccccccccc}
\hline
[Fe/H] & \multicolumn{18}{c}{$g-z$} \\
\hline
{} &
\multicolumn{2}{c}{t = 10.0 (Gyr)} &
\multicolumn{2}{c}{10.5} &
\multicolumn{2}{c}{11.0} &
\multicolumn{2}{c}{11.5} &
\multicolumn{2}{c}{12.0} &
\multicolumn{2}{c}{12.5} &
\multicolumn{2}{c}{13.0} &
\multicolumn{2}{c}{13.5} &
\multicolumn{2}{c}{14.0} \\
\hline
& $w/o$ & $w$ & $w/o$ & $w$ & $w/o$ & $w$ & $w/o$ & $w$ & $w/o$ & $w$ & $w/o$ & $w$ & $w/o$ & $w $& $w/o$ & $w$ & $w/o$ & $w$ \\
\hline
--2.5 & 0.730 & 0.631 & 0.744 & 0.652 & 0.758 & 0.668 & 0.771 & 0.696 & 0.783 & 0.723 & 0.793 & 0.745 & 0.803 & 0.766 & 0.811 & 0.779 & 0.820 & 0.792 \\
--2.4 & 0.741 & 0.647 & 0.755 & 0.664 & 0.769 & 0.678 & 0.781 & 0.705 & 0.794 & 0.731 & 0.804 & 0.753 & 0.814 & 0.774 & 0.823 & 0.789 & 0.831 & 0.803 \\
--2.3 & 0.752 & 0.664 & 0.766 & 0.677 & 0.781 & 0.689 & 0.793 & 0.714 & 0.805 & 0.738 & 0.815 & 0.761 & 0.826 & 0.783 & 0.835 & 0.799 & 0.844 & 0.815 \\
--2.2 & 0.764 & 0.682 & 0.779 & 0.691 & 0.793 & 0.700 & 0.805 & 0.723 & 0.817 & 0.746 & 0.828 & 0.769 & 0.838 & 0.793 & 0.847 & 0.810 & 0.857 & 0.828 \\
--2.1 & 0.778 & 0.702 & 0.792 & 0.706 & 0.806 & 0.712 & 0.818 & 0.733 & 0.830 & 0.755 & 0.841 & 0.779 & 0.852 & 0.802 & 0.861 & 0.821 & 0.871 & 0.841 \\
--2.0 & 0.792 & 0.722 & 0.806 & 0.722 & 0.820 & 0.725 & 0.833 & 0.745 & 0.845 & 0.764 & 0.855 & 0.789 & 0.866 & 0.813 & 0.876 & 0.833 & 0.886 & 0.854 \\
--1.9 & 0.808 & 0.744 & 0.822 & 0.740 & 0.836 & 0.740 & 0.848 & 0.757 & 0.860 & 0.774 & 0.871 & 0.799 & 0.882 & 0.824 & 0.892 & 0.846 & 0.903 & 0.868 \\
--1.8 & 0.825 & 0.767 & 0.839 & 0.760 & 0.853 & 0.756 & 0.865 & 0.771 & 0.877 & 0.785 & 0.888 & 0.811 & 0.899 & 0.837 & 0.910 & 0.860 & 0.921 & 0.883 \\
--1.7 & 0.845 & 0.793 & 0.858 & 0.783 & 0.872 & 0.775 & 0.884 & 0.786 & 0.896 & 0.798 & 0.908 & 0.824 & 0.919 & 0.850 & 0.930 & 0.874 & 0.941 & 0.899 \\
--1.6 & 0.866 & 0.821 & 0.880 & 0.808 & 0.893 & 0.797 & 0.906 & 0.804 & 0.918 & 0.811 & 0.929 & 0.839 & 0.940 & 0.865 & 0.951 & 0.891 & 0.963 & 0.916 \\
--1.5 & 0.891 & 0.851 & 0.905 & 0.838 & 0.918 & 0.824 & 0.930 & 0.825 & 0.942 & 0.827 & 0.953 & 0.856 & 0.965 & 0.882 & 0.976 & 0.909 & 0.987 & 0.934 \\
--1.4 & 0.920 & 0.884 & 0.933 & 0.873 & 0.946 & 0.857 & 0.958 & 0.851 & 0.971 & 0.847 & 0.982 & 0.875 & 0.993 & 0.901 & 1.004 & 0.928 & 1.016 & 0.953 \\
--1.3 & 0.953 & 0.921 & 0.967 & 0.913 & 0.980 & 0.900 & 0.992 & 0.885 & 1.004 & 0.872 & 1.015 & 0.899 & 1.026 & 0.924 & 1.037 & 0.950 & 1.048 & 0.974 \\
--1.2 & 0.992 & 0.960 & 1.005 & 0.959 & 1.019 & 0.954 & 1.031 & 0.932 & 1.043 & 0.908 & 1.054 & 0.930 & 1.065 & 0.951 & 1.075 & 0.976 & 1.084 & 0.998 \\
--1.1 & 1.035 & 1.003 & 1.050 & 1.010 & 1.065 & 1.017 & 1.077 & 0.996 & 1.090 & 0.963 & 1.100 & 0.973 & 1.110 & 0.984 & 1.118 & 1.006 & 1.126 & 1.025 \\
--1.0 & 1.083 & 1.049 & 1.100 & 1.062 & 1.116 & 1.079 & 1.129 & 1.071 & 1.142 & 1.057 & 1.152 & 1.036 & 1.161 & 1.029 & 1.167 & 1.042 & 1.173 & 1.055 \\
--0.9 & 1.134 & 1.098 & 1.152 & 1.115 & 1.170 & 1.134 & 1.184 & 1.138 & 1.198 & 1.144 & 1.207 & 1.118 & 1.216 & 1.092 & 1.220 & 1.088 & 1.224 & 1.092 \\
--0.8 & 1.184 & 1.147 & 1.203 & 1.166 & 1.222 & 1.185 & 1.237 & 1.194 & 1.252 & 1.207 & 1.261 & 1.194 & 1.271 & 1.169 & 1.274 & 1.147 & 1.278 & 1.136 \\
--0.7 & 1.232 & 1.197 & 1.251 & 1.215 & 1.271 & 1.232 & 1.287 & 1.243 & 1.302 & 1.257 & 1.313 & 1.255 & 1.323 & 1.245 & 1.328 & 1.220 & 1.332 & 1.195 \\
--0.6 & 1.278 & 1.247 & 1.297 & 1.262 & 1.317 & 1.276 & 1.333 & 1.289 & 1.349 & 1.302 & 1.361 & 1.308 & 1.373 & 1.311 & 1.379 & 1.297 & 1.386 & 1.271 \\
--0.5 & 1.321 & 1.294 & 1.340 & 1.307 & 1.359 & 1.318 & 1.375 & 1.332 & 1.392 & 1.344 & 1.405 & 1.356 & 1.419 & 1.367 & 1.429 & 1.365 & 1.438 & 1.356 \\
--0.4 & 1.362 & 1.340 & 1.380 & 1.351 & 1.398 & 1.360 & 1.414 & 1.374 & 1.431 & 1.385 & 1.446 & 1.400 & 1.461 & 1.417 & 1.475 & 1.424 & 1.488 & 1.431 \\
--0.3 & 1.401 & 1.384 & 1.418 & 1.393 & 1.434 & 1.402 & 1.451 & 1.415 & 1.468 & 1.426 & 1.484 & 1.443 & 1.501 & 1.462 & 1.518 & 1.475 & 1.534 & 1.489 \\
--0.2 & 1.438 & 1.425 & 1.453 & 1.435 & 1.469 & 1.444 & 1.486 & 1.457 & 1.502 & 1.468 & 1.520 & 1.485 & 1.538 & 1.504 & 1.558 & 1.521 & 1.577 & 1.538 \\
--0.1 & 1.472 & 1.465 & 1.487 & 1.475 & 1.501 & 1.486 & 1.518 & 1.499 & 1.535 & 1.510 & 1.554 & 1.527 & 1.573 & 1.544 & 1.595 & 1.562 & 1.617 & 1.581 \\
 0.0 & 1.505 & 1.502 & 1.518 & 1.514 & 1.532 & 1.527 & 1.549 & 1.539 & 1.566 & 1.552 & 1.586 & 1.567 & 1.606 & 1.582 & 1.629 & 1.601 & 1.653 & 1.619 \\
 0.1 & 1.534 & 1.537 & 1.548 & 1.551 & 1.562 & 1.565 & 1.579 & 1.578 & 1.596 & 1.593 & 1.616 & 1.606 & 1.637 & 1.619 & 1.661 & 1.637 & 1.685 & 1.655 \\
 0.2 & 1.562 & 1.571 & 1.576 & 1.585 & 1.590 & 1.600 & 1.607 & 1.614 & 1.624 & 1.629 & 1.644 & 1.642 & 1.665 & 1.654 & 1.689 & 1.671 & 1.714 & 1.689 \\
 0.3 & 1.587 & 1.603 & 1.602 & 1.617 & 1.616 & 1.631 & 1.633 & 1.645 & 1.650 & 1.661 & 1.671 & 1.674 & 1.691 & 1.687 & 1.716 & 1.705 & 1.740 & 1.722 \\
 0.4 & 1.611 & 1.633 & 1.626 & 1.646 & 1.641 & 1.659 & 1.658 & 1.673 & 1.675 & 1.688 & 1.696 & 1.704 & 1.716 & 1.719 & 1.740 & 1.737 & 1.764 & 1.755 \\
 0.5 & 1.633 & 1.662 & 1.648 & 1.673 & 1.664 & 1.683 & 1.682 & 1.697 & 1.699 & 1.711 & 1.719 & 1.729 & 1.739 & 1.749 & 1.762 & 1.769 & 1.785 & 1.790 \\
\hline
\end{tabular}
\tablecomments{The entire model data are available at http://web.yonsei.ac.kr/cosmic/data/YEPS.htm.}.
\end{table}
\end{rotate}

\clearpage
\begin{rotate}
\begin{table}
\tiny
\caption{\footnotesize Theoretical $u-z$ colors for synthetic GCs with various ages ($t$) based on the YEPS models without ($w/o$) and with ($w$) the inclusion of the HB prescriptions (the 5th-order polynomial fits to the model data).\label{tbl-2}}
\begin{tabular}{ccccccccccccccccccc}
\hline
[Fe/H] &
\multicolumn{18}{c}{$u-z$} \\
\hline
{} &
\multicolumn{2}{c}{t = 10.0 (Gyr)} &
\multicolumn{2}{c}{10.5} &
\multicolumn{2}{c}{11.0} &
\multicolumn{2}{c}{11.5} &
\multicolumn{2}{c}{12.0} &
\multicolumn{2}{c}{12.5} &
\multicolumn{2}{c}{13.0} &
\multicolumn{2}{c}{13.5} &
\multicolumn{2}{c}{14.0} \\
\hline
& $w/o$ & $w$ & $w/o$ & $w$ & $w/o$ & $w$ & $w/o$ & $w$ & $w/o$ & $w$ & $w/o$ & $w$ & $w/o$ & $w $& $w/o$ & $w$ & $w/o$ & $w$ \\
\hline
--2.5 & 2.114 & 2.025 & 2.131 & 2.026 & 2.149 & 2.026 & 2.164 & 2.033 & 2.179 & 2.041 & 2.192 & 2.059 & 2.206 & 2.078 & 2.218 & 2.099 & 2.230 & 2.120 \\
--2.4 & 2.136 & 2.050 & 2.154 & 2.050 & 2.172 & 2.050 & 2.187 & 2.056 & 2.203 & 2.062 & 2.217 & 2.080 & 2.232 & 2.098 & 2.245 & 2.121 & 2.258 & 2.143 \\
--2.3 & 2.159 & 2.076 & 2.178 & 2.077 & 2.196 & 2.077 & 2.212 & 2.081 & 2.229 & 2.084 & 2.244 & 2.102 & 2.259 & 2.120 & 2.273 & 2.144 & 2.287 & 2.167 \\
--2.2 & 2.184 & 2.104 & 2.203 & 2.105 & 2.222 & 2.105 & 2.239 & 2.107 & 2.256 & 2.108 & 2.272 & 2.126 & 2.288 & 2.143 & 2.303 & 2.168 & 2.318 & 2.193 \\
--2.1 & 2.211 & 2.134 & 2.230 & 2.135 & 2.250 & 2.135 & 2.268 & 2.135 & 2.286 & 2.134 & 2.302 & 2.151 & 2.319 & 2.167 & 2.335 & 2.193 & 2.351 & 2.219 \\
--2.0 & 2.240 & 2.167 & 2.260 & 2.167 & 2.280 & 2.168 & 2.299 & 2.165 & 2.317 & 2.163 & 2.335 & 2.178 & 2.352 & 2.193 & 2.369 & 2.221 & 2.386 & 2.248 \\
--1.9 & 2.271 & 2.203 & 2.292 & 2.203 & 2.313 & 2.204 & 2.332 & 2.198 & 2.351 & 2.193 & 2.370 & 2.208 & 2.388 & 2.222 & 2.405 & 2.250 & 2.423 & 2.277 \\
--1.8 & 2.306 & 2.242 & 2.327 & 2.242 & 2.348 & 2.242 & 2.368 & 2.234 & 2.388 & 2.227 & 2.407 & 2.240 & 2.426 & 2.253 & 2.444 & 2.281 & 2.462 & 2.309 \\
--1.7 & 2.343 & 2.284 & 2.365 & 2.284 & 2.387 & 2.285 & 2.408 & 2.274 & 2.428 & 2.265 & 2.448 & 2.276 & 2.467 & 2.287 & 2.486 & 2.315 & 2.504 & 2.342 \\
--1.6 & 2.384 & 2.331 & 2.407 & 2.331 & 2.430 & 2.332 & 2.451 & 2.319 & 2.472 & 2.307 & 2.492 & 2.316 & 2.512 & 2.325 & 2.531 & 2.352 & 2.549 & 2.379 \\
--1.5 & 2.429 & 2.382 & 2.453 & 2.383 & 2.476 & 2.384 & 2.498 & 2.369 & 2.520 & 2.355 & 2.540 & 2.361 & 2.560 & 2.367 & 2.579 & 2.393 & 2.597 & 2.418 \\
--1.4 & 2.479 & 2.439 & 2.504 & 2.440 & 2.528 & 2.441 & 2.551 & 2.426 & 2.573 & 2.411 & 2.593 & 2.413 & 2.613 & 2.416 & 2.631 & 2.439 & 2.649 & 2.460 \\
--1.3 & 2.534 & 2.502 & 2.560 & 2.504 & 2.586 & 2.506 & 2.609 & 2.491 & 2.632 & 2.476 & 2.651 & 2.473 & 2.671 & 2.472 & 2.689 & 2.490 & 2.706 & 2.507 \\
--1.2 & 2.594 & 2.570 & 2.621 & 2.574 & 2.648 & 2.577 & 2.672 & 2.565 & 2.695 & 2.551 & 2.715 & 2.544 & 2.735 & 2.538 & 2.751 & 2.548 & 2.767 & 2.559 \\
--1.1 & 2.659 & 2.643 & 2.687 & 2.650 & 2.716 & 2.656 & 2.740 & 2.648 & 2.764 & 2.639 & 2.784 & 2.626 & 2.803 & 2.616 & 2.818 & 2.615 & 2.833 & 2.618 \\
--1.0 & 2.728 & 2.720 & 2.758 & 2.730 & 2.788 & 2.740 & 2.813 & 2.739 & 2.839 & 2.738 & 2.858 & 2.723 & 2.877 & 2.708 & 2.891 & 2.693 & 2.904 & 2.685 \\
--0.9 & 2.800 & 2.800 & 2.832 & 2.815 & 2.864 & 2.829 & 2.890 & 2.835 & 2.917 & 2.843 & 2.937 & 2.830 & 2.957 & 2.815 & 2.969 & 2.785 & 2.982 & 2.764 \\
--0.8 & 2.875 & 2.883 & 2.908 & 2.901 & 2.942 & 2.920 & 2.970 & 2.934 & 2.998 & 2.949 & 3.019 & 2.942 & 3.040 & 2.933 & 3.053 & 2.893 & 3.066 & 2.857 \\
--0.7 & 2.952 & 2.966 & 2.987 & 2.989 & 3.022 & 3.013 & 3.052 & 3.032 & 3.081 & 3.052 & 3.104 & 3.052 & 3.127 & 3.052 & 3.141 & 3.015 & 3.155 & 2.971 \\
--0.6 & 3.030 & 3.051 & 3.066 & 3.078 & 3.102 & 3.106 & 3.134 & 3.128 & 3.166 & 3.152 & 3.191 & 3.160 & 3.216 & 3.168 & 3.234 & 3.144 & 3.250 & 3.108 \\
--0.5 & 3.109 & 3.137 & 3.146 & 3.168 & 3.183 & 3.199 & 3.217 & 3.223 & 3.251 & 3.248 & 3.279 & 3.264 & 3.307 & 3.280 & 3.329 & 3.274 & 3.349 & 3.259 \\
--0.4 & 3.190 & 3.225 & 3.228 & 3.258 & 3.266 & 3.291 & 3.301 & 3.318 & 3.336 & 3.343 & 3.368 & 3.365 & 3.399 & 3.387 & 3.425 & 3.399 & 3.451 & 3.410 \\
--0.3 & 3.274 & 3.316 & 3.311 & 3.350 & 3.349 & 3.384 & 3.385 & 3.411 & 3.422 & 3.438 & 3.456 & 3.464 & 3.491 & 3.492 & 3.523 & 3.517 & 3.554 & 3.547 \\
--0.2 & 3.360 & 3.409 & 3.397 & 3.444 & 3.433 & 3.477 & 3.471 & 3.505 & 3.508 & 3.533 & 3.546 & 3.563 & 3.583 & 3.594 & 3.620 & 3.629 & 3.657 & 3.669 \\
--0.1 & 3.450 & 3.506 & 3.485 & 3.539 & 3.520 & 3.571 & 3.558 & 3.600 & 3.595 & 3.629 & 3.635 & 3.662 & 3.674 & 3.695 & 3.715 & 3.736 & 3.757 & 3.778 \\
 0.0 & 3.542 & 3.606 & 3.575 & 3.636 & 3.609 & 3.665 & 3.645 & 3.696 & 3.682 & 3.726 & 3.723 & 3.761 & 3.764 & 3.795 & 3.809 & 3.836 & 3.854 & 3.878 \\
 0.1 & 3.635 & 3.705 & 3.666 & 3.732 & 3.698 & 3.759 & 3.733 & 3.791 & 3.770 & 3.823 & 3.811 & 3.858 & 3.852 & 3.893 & 3.899 & 3.932 & 3.946 & 3.970 \\
 0.2 & 3.726 & 3.801 & 3.755 & 3.826 & 3.785 & 3.851 & 3.820 & 3.884 & 3.855 & 3.918 & 3.897 & 3.952 & 3.938 & 3.987 & 3.986 & 4.023 & 4.033 & 4.058 \\
 0.3 & 3.808 & 3.887 & 3.838 & 3.913 & 3.868 & 3.939 & 3.903 & 3.972 & 3.938 & 4.005 & 3.979 & 4.040 & 4.020 & 4.074 & 4.067 & 4.109 & 4.113 & 4.143 \\
 0.4 & 3.881 & 3.964 & 3.913 & 3.992 & 3.945 & 4.021 & 3.981 & 4.052 & 4.016 & 4.084 & 4.056 & 4.119 & 4.097 & 4.153 & 4.142 & 4.190 & 4.187 & 4.226 \\
 0.5 & 3.943 & 4.030 & 3.979 & 4.063 & 4.014 & 4.096 & 4.052 & 4.125 & 4.089 & 4.154 & 4.129 & 4.189 & 4.169 & 4.224 & 4.213 & 4.266 & 4.257 & 4.309 \\
\hline
\end{tabular}
\tablecomments{The entire model data are available at http://web.yonsei.ac.kr/cosmic/data/YEPS.htm.}.
\end{table}
\end{rotate}

\clearpage
\begin{rotate}
\begin{table}
\tiny
\caption{\footnotesize Theoretical $u-g$ colors for synthetic GCs with various ages ($t$) based on the YEPS models without ($w/o$) and with ($w$) the inclusion of the HB prescriptions (the 5th-order polynomial fits to the model data).\label{tbl-2}}
\begin{tabular}{ccccccccccccccccccc}
\hline
[Fe/H] &
\multicolumn{18}{c}{$u-g$} \\
\hline
{} &
\multicolumn{2}{c}{t = 10.0 (Gyr)} &
\multicolumn{2}{c}{10.5} &
\multicolumn{2}{c}{11.0} &
\multicolumn{2}{c}{11.5} &
\multicolumn{2}{c}{12.0} &
\multicolumn{2}{c}{12.5} &
\multicolumn{2}{c}{13.0} &
\multicolumn{2}{c}{13.5} &
\multicolumn{2}{c}{14.0} \\
\hline
& $w/o$ & $w$ & $w/o$ & $w$ & $w/o$ & $w$ & $w/o$ & $w$ & $w/o$ & $w$ & $w/o$ & $w$ & $w/o$ & $w $& $w/o$ & $w$ & $w/o$ & $w$ \\
\hline
--2.5 & 1.380 & 1.386 & 1.383 & 1.366 & 1.386 & 1.342 & 1.388 & 1.326 & 1.391 & 1.308 & 1.394 & 1.310 & 1.398 & 1.311 & 1.402 & 1.318 & 1.406 & 1.327 \\
--2.4 & 1.393 & 1.397 & 1.397 & 1.381 & 1.400 & 1.366 & 1.404 & 1.345 & 1.407 & 1.326 & 1.411 & 1.325 & 1.415 & 1.323 & 1.420 & 1.331 & 1.425 & 1.339 \\
--2.3 & 1.407 & 1.408 & 1.411 & 1.397 & 1.415 & 1.387 & 1.419 & 1.365 & 1.424 & 1.344 & 1.429 & 1.340 & 1.433 & 1.336 & 1.439 & 1.344 & 1.444 & 1.352 \\
--2.2 & 1.421 & 1.421 & 1.426 & 1.413 & 1.430 & 1.408 & 1.436 & 1.385 & 1.441 & 1.363 & 1.446 & 1.356 & 1.452 & 1.350 & 1.458 & 1.357 & 1.464 & 1.365 \\
--2.1 & 1.435 & 1.434 & 1.441 & 1.430 & 1.446 & 1.428 & 1.452 & 1.406 & 1.458 & 1.383 & 1.464 & 1.373 & 1.470 & 1.364 & 1.477 & 1.372 & 1.483 & 1.379 \\
--2.0 & 1.450 & 1.448 & 1.457 & 1.447 & 1.463 & 1.449 & 1.470 & 1.427 & 1.476 & 1.404 & 1.483 & 1.392 & 1.490 & 1.380 & 1.496 & 1.388 & 1.503 & 1.394 \\
--1.9 & 1.466 & 1.463 & 1.473 & 1.465 & 1.480 & 1.469 & 1.488 & 1.448 & 1.495 & 1.426 & 1.502 & 1.412 & 1.509 & 1.398 & 1.516 & 1.404 & 1.523 & 1.410 \\
--1.8 & 1.483 & 1.479 & 1.490 & 1.484 & 1.498 & 1.489 & 1.506 & 1.470 & 1.514 & 1.449 & 1.521 & 1.433 & 1.529 & 1.417 & 1.536 & 1.423 & 1.543 & 1.427 \\
--1.7 & 1.500 & 1.496 & 1.508 & 1.503 & 1.517 & 1.510 & 1.525 & 1.493 & 1.534 & 1.474 & 1.542 & 1.456 & 1.550 & 1.438 & 1.557 & 1.442 & 1.564 & 1.445 \\
--1.6 & 1.518 & 1.515 & 1.527 & 1.524 & 1.536 & 1.531 & 1.545 & 1.516 & 1.554 & 1.500 & 1.562 & 1.481 & 1.571 & 1.461 & 1.578 & 1.463 & 1.585 & 1.464 \\
--1.5 & 1.536 & 1.535 & 1.546 & 1.545 & 1.556 & 1.553 & 1.566 & 1.540 & 1.575 & 1.527 & 1.584 & 1.508 & 1.593 & 1.487 & 1.600 & 1.487 & 1.607 & 1.485 \\
--1.4 & 1.556 & 1.557 & 1.567 & 1.568 & 1.577 & 1.575 & 1.588 & 1.566 & 1.598 & 1.556 & 1.606 & 1.537 & 1.615 & 1.516 & 1.623 & 1.512 & 1.630 & 1.508 \\
--1.3 & 1.576 & 1.582 & 1.588 & 1.591 & 1.599 & 1.599 & 1.610 & 1.592 & 1.621 & 1.586 & 1.630 & 1.569 & 1.639 & 1.549 & 1.646 & 1.541 & 1.654 & 1.533 \\
--1.2 & 1.598 & 1.608 & 1.610 & 1.616 & 1.623 & 1.623 & 1.634 & 1.620 & 1.645 & 1.618 & 1.654 & 1.603 & 1.664 & 1.585 & 1.671 & 1.573 & 1.678 & 1.562 \\
--1.1 & 1.620 & 1.636 & 1.634 & 1.643 & 1.647 & 1.649 & 1.659 & 1.650 & 1.670 & 1.652 & 1.680 & 1.639 & 1.690 & 1.625 & 1.697 & 1.608 & 1.705 & 1.593 \\
--1.0 & 1.644 & 1.667 & 1.658 & 1.672 & 1.672 & 1.677 & 1.685 & 1.681 & 1.697 & 1.687 & 1.707 & 1.678 & 1.718 & 1.669 & 1.725 & 1.648 & 1.732 & 1.629 \\
--0.9 & 1.669 & 1.699 & 1.684 & 1.703 & 1.699 & 1.707 & 1.712 & 1.715 & 1.725 & 1.725 & 1.736 & 1.720 & 1.747 & 1.717 & 1.754 & 1.693 & 1.762 & 1.670 \\
--0.8 & 1.696 & 1.733 & 1.712 & 1.736 & 1.728 & 1.740 & 1.742 & 1.751 & 1.756 & 1.764 & 1.767 & 1.763 & 1.778 & 1.766 & 1.786 & 1.743 & 1.794 & 1.718 \\
--0.7 & 1.725 & 1.769 & 1.742 & 1.772 & 1.758 & 1.777 & 1.773 & 1.791 & 1.788 & 1.806 & 1.800 & 1.809 & 1.812 & 1.816 & 1.821 & 1.797 & 1.830 & 1.774 \\
--0.6 & 1.757 & 1.807 & 1.774 & 1.812 & 1.791 & 1.818 & 1.807 & 1.834 & 1.823 & 1.850 & 1.836 & 1.857 & 1.849 & 1.868 & 1.859 & 1.855 & 1.868 & 1.838 \\
--0.5 & 1.791 & 1.847 & 1.809 & 1.855 & 1.827 & 1.865 & 1.844 & 1.881 & 1.861 & 1.897 & 1.875 & 1.907 & 1.889 & 1.919 & 1.901 & 1.916 & 1.912 & 1.910 \\
--0.4 & 1.828 & 1.889 & 1.847 & 1.903 & 1.866 & 1.918 & 1.884 & 1.933 & 1.902 & 1.947 & 1.918 & 1.959 & 1.934 & 1.972 & 1.947 & 1.978 & 1.960 & 1.984 \\
--0.3 & 1.869 & 1.934 & 1.889 & 1.955 & 1.909 & 1.976 & 1.928 & 1.988 & 1.947 & 2.000 & 1.965 & 2.014 & 1.983 & 2.027 & 1.999 & 2.041 & 2.014 & 2.058 \\
--0.2 & 1.916 & 1.982 & 1.937 & 2.011 & 1.957 & 2.037 & 1.977 & 2.047 & 1.998 & 2.056 & 2.018 & 2.071 & 2.037 & 2.084 & 2.056 & 2.105 & 2.075 & 2.128 \\
--0.1 & 1.971 & 2.037 & 1.991 & 2.069 & 2.012 & 2.095 & 2.033 & 2.106 & 2.054 & 2.115 & 2.075 & 2.131 & 2.096 & 2.144 & 2.118 & 2.169 & 2.139 & 2.194 \\
 0.0 & 2.034 & 2.098 & 2.053 & 2.128 & 2.073 & 2.151 & 2.094 & 2.164 & 2.114 & 2.177 & 2.136 & 2.193 & 2.158 & 2.208 & 2.180 & 2.233 & 2.203 & 2.256 \\
 0.1 & 2.102 & 2.166 & 2.120 & 2.185 & 2.138 & 2.203 & 2.158 & 2.220 & 2.177 & 2.238 & 2.198 & 2.256 & 2.219 & 2.273 & 2.242 & 2.295 & 2.264 & 2.315 \\
 0.2 & 2.169 & 2.233 & 2.186 & 2.241 & 2.203 & 2.252 & 2.220 & 2.274 & 2.238 & 2.296 & 2.258 & 2.315 & 2.278 & 2.336 & 2.299 & 2.353 & 2.321 & 2.370 \\
 0.3 & 2.226 & 2.289 & 2.242 & 2.294 & 2.259 & 2.302 & 2.276 & 2.326 & 2.294 & 2.349 & 2.313 & 2.369 & 2.332 & 2.390 & 2.353 & 2.406 & 2.374 & 2.423 \\
 0.4 & 2.270 & 2.332 & 2.288 & 2.344 & 2.305 & 2.356 & 2.324 & 2.377 & 2.342 & 2.396 & 2.361 & 2.415 & 2.381 & 2.434 & 2.402 & 2.453 & 2.423 & 2.472 \\
 0.5 & 2.306 & 2.365 & 2.325 & 2.392 & 2.344 & 2.421 & 2.364 & 2.427 & 2.384 & 2.438 & 2.405 & 2.456 & 2.426 & 2.472 & 2.448 & 2.495 & 2.471 & 2.519 \\
\hline
\end{tabular}
\tablecomments{The entire model data are available at http://web.yonsei.ac.kr/cosmic/data/YEPS.htm.}.
\end{table}
\end{rotate}

\clearpage
\begin{table}
\scriptsize
\begin{center}
\caption{The $u$-, $g$-, and $z$-band mags and their observational errors for the M87 GCs \label{tbl-5}}
\begin{tabular}{c c c c c c c c c}
\hline
GC ID & RA (J2000) & DEC (J2000) & $u_{0}$ & u error & $g_{0}$\tablenotemark{a} & g error\tablenotemark{a} & $z_{0}\tablenotemark{a} $ & z error\tablenotemark{a}  \\
            &                        &                       & {\tiny (WFPC3 F336W)} &               & {\tiny (ACS/WFC F475W)}   & & {\tiny (ACS/WFC F850LP)} \\              
\hline
  1 & 187.7056427 & 12.3909702 & 25.375 & 0.212 & 23.275 & 0.083 & 22.668 & 0.090 \\
  2 & 187.7063599 & 12.3914337 & 25.829 & 0.275 & 23.817 & 0.108 & 22.580 & 0.119 \\
  3 & 187.7055817 & 12.3919382 & 24.102 & 0.040 & 22.122 & 0.027 & 20.749 & 0.037 \\
  4 & 187.7067566 & 12.3914499 & 26.307 & 0.230 & 24.046 & 0.112 & 22.408 & 0.110 \\
  5 & 187.7069092 & 12.3912258 & 25.570 & 0.115 & 23.810 & 0.093 & 22.259 & 0.078 \\
  6 & 187.7063293 & 12.3921633 & 25.516 & 0.142 & 23.002 & 0.031 & 21.521 & 0.042 \\
  7 & 187.7062683 & 12.3902464 & 25.836 & 0.202 & 23.370 & 0.043 & 21.957 & 0.046 \\
  8 & 187.7059631 & 12.3899851 & 24.170 & 0.046 & 21.829 & 0.022 & 20.451 & 0.013 \\
  9 & 187.7071381 & 12.3918676 & 23.499 & 0.021 & 21.169 & 0.015 & 19.779 & 0.017 \\
 10 & 187.7046356 & 12.3907633 & 26.113 & 0.226 & 24.061 & 0.100 & 22.583 & 0.057 \\
\nodata & \nodata & \nodata & \nodata & \nodata & \nodata & \nodata & \nodata & \nodata \\
\nodata & \nodata & \nodata & \nodata & \nodata & \nodata & \nodata & \nodata & \nodata \\
\nodata & \nodata & \nodata & \nodata & \nodata & \nodata & \nodata & \nodata & \nodata \\
\hline
\end{tabular}
\tablenotetext{a}{The $g$- and $z$-band data are obtained from \citet{jordan09}.}
\tablecomments{A sample table is presented here; Table 5 is available in its entirety in the electronic version of the article.}
\end{center}
\end{table}

\clearpage
\begin{table}
\small
\begin{center}
\caption{The median photometric errors in $g-z$, $u-z$, and $u-g$ of five magnitude bins for the M87 GCs \label{tbl-5}}
\begin{tabular}{c c c c c}
\hline
\hline
Mag bins & Number of GCs & {\,\,\,\,\,\,\,\,\,\,$g-z$ error} & {\,\,\,\,\,\,\,\,\,\,$u-z$ error} & {\,\,\,\,\,\,\,\,\,\,$u-g$ error}  \\
\hline
            $u_0$ $\le$ 23.0  &  20  &  {\,\,\,\,\,\,\,\,\,\,0.019}  &   {\,\,\,\,\,\,\,\,\,\,0.018}   &  {\,\,\,\,\,\,\,\,\,\,0.024} \\
23.0 $<$ $u_0$ $\le$ 24.0  &  87  &     {\,\,\,\,\,\,\,\,\,\,0.022}  &   {\,\,\,\,\,\,\,\,\,\,0.026}   &  {\,\,\,\,\,\,\,\,\,\,0.028} \\
24.0 $<$ $u_0$ $\le$ 25.0  &  172  &     {\,\,\,\,\,\,\,\,\,\,0.030}  &   {\,\,\,\,\,\,\,\,\,\,0.054}   &  {\,\,\,\,\,\,\,\,\,\,0.054} \\
25.0 $<$ $u_0$ $\le$ 26.0  &  247  &     {\,\,\,\,\,\,\,\,\,\,0.043}  &   {\,\,\,\,\,\,\,\,\,\,0.111}   &  {\,\,\,\,\,\,\,\,\,\,0.110} \\
            $u_0$ $>$  26.0                  & 65     &     {\,\,\,\,\,\,\,\,\,\,0.058}  & {\,\,\,\,\,\,\,\,\,\,0.175}  &  {\,\,\,\,\,\,\,\,\,\,0.175}    \\
\hline
Entire Sample & 591   &   {\,\,\,\,\,\,\,\,\,\,0.035}   &  {\,\,\,\,\,\,\,\,\,\,0.080}    &  {\,\,\,\,\,\,\,\,\,\,0.078}  \\
\hline
\end{tabular}
\end{center}
\end{table}

\end{document}